# Novel gas target for Laser Wakefield Accelerators

C. Aniculaesei,[1,a] Hyung Taek Kim,[1,2,a] Byeong Ju Yoo,[1] Kyung Hwan Oh[1] and Chang Hee Nam[1,3]

[1]*Center for Relativistic Laser Science, Institute for Basic Science (IBS), Gwangju 61005, Republic of Korea.*

[2]*Advanced Photonics Research Institute, Gwangju Institute of Science and Technology(GIST), Gwangju 61005, Republic of Korea*

[3]*Department of Physics and Photon Science, GIST, Gwangju 61005, Republic of Korea*

## Abstract

A novel gas target for interactions between high power lasers and gaseous medium, especially for laser wakefield accelerators, has been designed, manufactured and characterized. The gas target has been designed to provide a uniform density profile along the central gas cell axis by combining a gas cell and slit nozzle. The gas density can be tuned from $10^{16}$ atoms/cm$^3$ to $10^{19}$ atoms/cm$^3$ and the gas target length can be varied from 0 to 10 cm, both changes can be made simultaneously while keeping the flat-top gas profile. The gas distributions inside the gas cell have been measured using interferometry and validated using computational fluid dynamics.

## I. INTRODUCTION

The laser wakefield acceleration (LWFA)[1] uses nonlinear plasma waves[2], created by the interaction between a high power laser pulse and an under-dense plasma medium to trap and accelerate plasma electrons towards relativistic velocities. One of the important issues in LWFA research is the enhancement of accelerated electron energy. The energy of the accelerated electrons can be estimated by a formula[3], $\Delta E [\text{GeV}] \approx 1.7 \left(\frac{P[\text{TW}]}{100}\right)^{1/3} \left(\frac{10^{18}}{n_p[\text{cm}^{-3}]}\right)^{2/3} \left(\frac{0.8}{\lambda_0[\mu m]}\right)^{4/3}$, where $\Delta E$ is the energy gained by the electrons, $P$ is the laser power, $n_p$ is the plasma density and $\lambda_0$ is the laser wavelength. In general, the maximum achievable electron energy increases as the acceleration length and the laser power increases and/or plasma density is lowered. Thus, a laser pulse below 100-TW peak power can produce a few 100 MeV's electron beams in millimeters long gas targets[4], while a PW laser pulse[5,6] can generate several GeV electron beams with centimeters long gas targets[7–9]. As the laser power increases rapidly to multi-PW level, a long and uniform gas medium over 10-cm length is required to obtain an electron beam over 10 GeV energy. However, conventional supersonic gas nozzles[10], conical[11] or slit

---

[a] Hyung Taek Kim: htkim@gist.ac.kr

Constantin Aniculaesei: CA182@ibs.re.kr

type[12], cannot generate such a long and uniform density profile[13] due to a high gas load and non-uniform gas expansions.

A long gas medium can be produced by using gas cells[14], which are small chambers filled with gas having apertures for the entrance of laser and the exit of accelerated electrons. Some gas cells can have a variable length from 0 up to a few centimeters[15]. One of the main flaws of such gas cells design is its length which has to be limited when a low density uniform profile is required. The required gas medium to generate 10 GeV electron beam with multi-PW lasers has to be uniform and tens-of-centimeter long at the density about $10^{17}$ atoms/cm$^3$, but gas cells cannot fulfill this requirement because the gas inlet creates a gas density spike with such a low gas density. As an example, we perform a computational fluid dynamics (CFD) simulation for a typical gas cell of 2 cm × 2 cm × 3 cm (height × width × length) with a 2-mm inlet and 2-mm 2 outlets, as shown in Fig. 1. The gas density spike is clearly visible around the inlet and the density difference between the spike and the uniform region is more than 30 % in this case. The gas density spike issue can be mitigated by reducing the backing pressure, reducing the inlet size or decreasing the gas cell length, but cannot be eliminated in a typical gas cell. The reduction in backing pressure and inlet size subsequently limits the controllable density range. Thus, a long uniform variable-length gas cell cannot allow a wide range of gas density control.

In this paper, we report a new type of gas medium, suitable for 10 GeV electron acceleration driven by multi-PW laser pulses[22], by combining elements from a typical gas cell and a slit nozzle to provide a gas target with a uniform gas density profile whose length and density are tunable. To avoid unnecessary repetition or confusion, in this article we call "*SlitCell*" the novel gas target. We have performed a series of CFD simulations to verify if the concept of *SlitCell* is realizable. Based on CFD simulation results we have designed and manufactured a *SlitCell* having variable length from 0 to 10 cm. The density distribution inside the *SlitCell* has been characterized through interferometry.

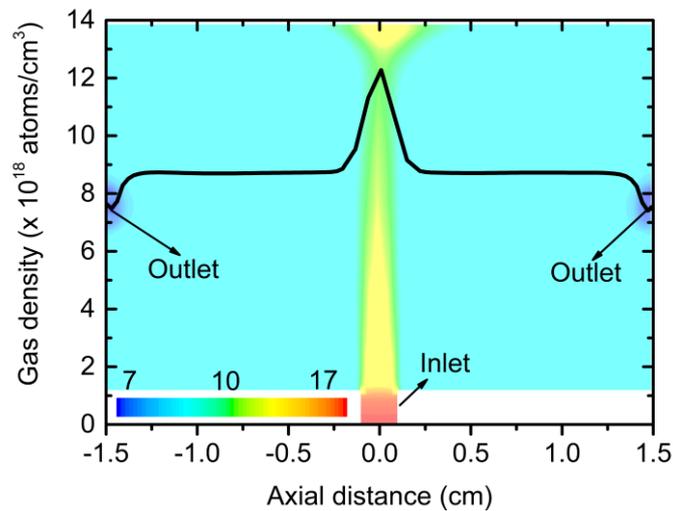

**Fig. 1 Simulated 2D contour density of a cross-section through the center of a 3D gas cell (background) and the gas density taken along a line through the center (foreground). The gas cell has 2 mm gas inlet, 2 cm height, 2 cm width, 3 cm length and the backing pressure is He at 1 bar. The gas density spike is present around the gas inlet. The colormap is in logarithmic scale in units of $10^{18}$ atoms/cm$^{-3}$.**

## II. DESIGN, SIMULATION AND CHARACTERIZATION

### 1. *SLITCELL* DESIGN

A suitable gas target for LWFA should have a tunable density and length, with a uniform density profile. To fulfill these criteria, especially for long acceleration lengths, we have decided to design a shaped inlet for the gas cells to eliminate the gas density spike mentioned in the previous section and obtain a uniform gas density for various ranges of gas density and medium lengths. After comprehensive simulations using the commercial CFD code (ANSYS Fluent), we have modified the inlet of gas cells to have a slit nozzle shape rather than the usual cylindrical inlet shape. As shown in Fig. 2, the slit nozzle part has a cylindrical gas inlet with a diameter of 0.1 mm and length of 2-mm, and a rectangular outlet at 20 mm above the inlet with the dimension of 20x20x90 mm$^3$. The gas inlet is fed with gas through a 2 mm diameter Quick-Connect type connector shown in Fig. 2 and Fig. 3. The slit nozzle outlet fits exactly one side of the gas cell, as shown in Fig. 3, and the gas flows through this surface to fill the gas cell section. The gas cell has five surfaces on top of the slit nozzle part to contain the gas medium; two gas outlets are placed at front and back sides of the cell, and three optical windows are mounted at two sides and top surface. The windows permit optical access to the interior of the gas cell for probing LWFA process and monitoring plasma radiations. The two ends of the gas cell have two 1.4 mm diameter outlets, one outlet placed at the front and the other outlet placed on a movable rectangular insert. The movable part is motorized and controlled via computer to change the effective distance between two outlets between 0 and 100 mm. The present "*SlitCell*" design has been characterized using computational fluid dynamics simulations which are presented in the next section.

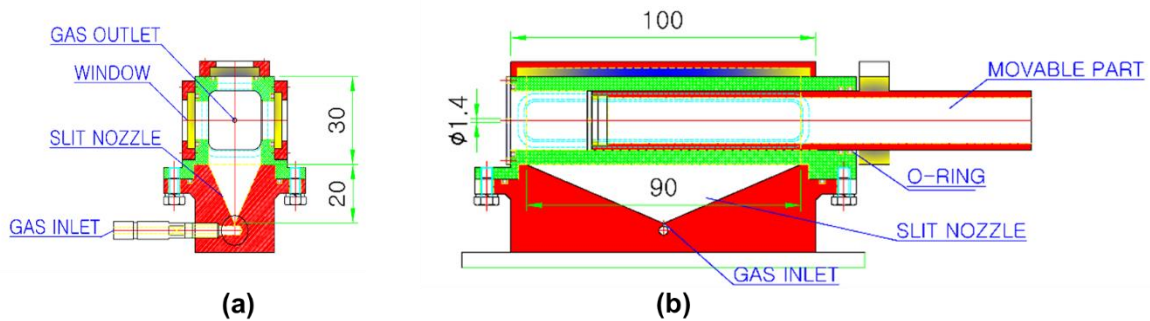

**Fig. 2 Drawing of the *SlitCell* for (a) the front view and (b) the side view. The movable part mounted on a translation stage to vary the distance between two outlets. The *Slitcell* is equipped with two lateral and one top fused silica windows for monitoring visible plasma radiations.**

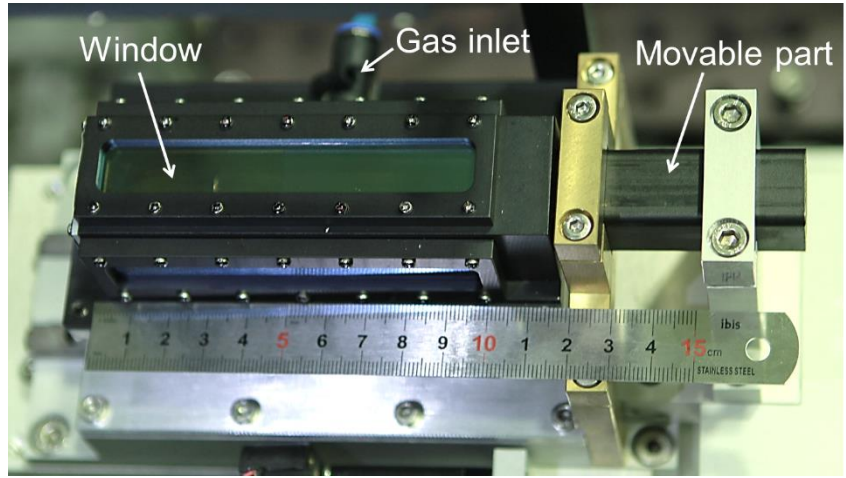

**Fig. 3.** Photograph of the *Slitcell*

## 2. CFD SIMULATIONS FOR SLITCELL DESIGN

We have performed CFD simulations using the ANSYS Fluent that can solve the Navier-Stokes equations[16,17] on a 3-dimensional mesh only containing hexahedral cells with various volumes between $10^{-12}$ m$^3$ and $10^{-16}$ m$^3$. On average, each mesh contained at least $3 \times 10^6$ cells. A turbulent model, the k-ω shear stress transport model[18], has been used with double accuracy. The gas inlet has been chosen as pressure inlet which feeds He gas of various pressures at 300-K temperature. The outlets are chosen as pressure outlets at $10^{-3}$ mbar pressure and the walls are adiabatic. Also mesh-independence tests have been run to ensure that the simulation result does not depend on the mesh size.

The ANSYS Fluent simulations have been run with two different objectives:
- to establish the most suitable geometry, dimensions and shape for final design of the *SlitCell* to provide the desired density profile.
- to predict the density profile of the *SlitCell* for different experimental conditions such as backing pressures or medium lengths.

The reasons why we have chosen this kind of geometry for the *SlitCell* are as follows. The gas enters the slit nozzle through a small 0.1 mm diameter inlet, as shown in Fig. , and expands continuously inside the slit nozzle as it propagates towards the nozzle outlet. The slit nozzle shape and length ensures that the gas density profile, as it exits the slit nozzle and enters the gas cell, is uniform. The gas fills the gas cell uniformly and exits through two 1.4 mm diameter outlets. Although the density profile initially presents a very strong density spike (see the black profile in Fig. ) as it enters the gas cell at 20 mm above the inlet, the difference between the density spike and flat profile is around 10%. The gas propagates further to the central axis of the cell part at 35 mm above the inlet and the density spike along the central axis reduces a negligible level of 2 % difference. These simulation results from ANSYS Fluent with Helium at pressure of 10 bar show that the combination of the slit nozzle and gas cell forms the proper geometry for the new gas target, the *SlitCell*.

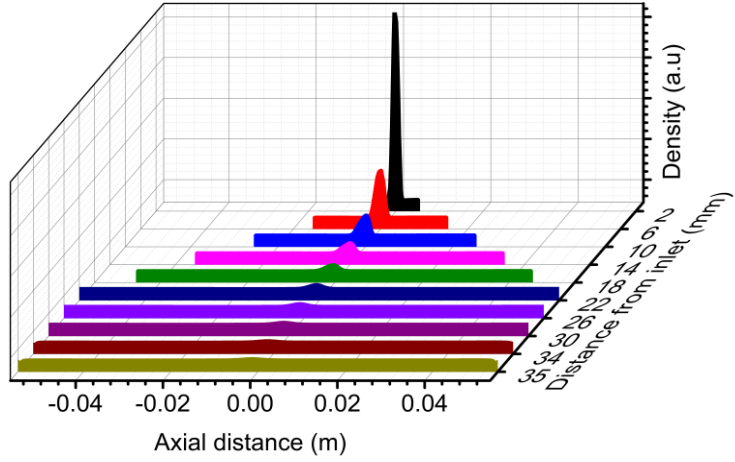

**Fig. 4 Lineouts taken along the central part of the *SlitCell* showing the evolution of the gas density profile as it propagates from the inlet to the gas cell section. The profiles are plotted in steps of 4 mm beginning with 2 mm above the inlet (black curve). The density profile presents a very strong density spike that decreases very fast as it propagates through the *SlitCell*.**

## 3. CHARACTERIZATION METHOD

The gas density profile of the *SlitCell* has been retrieved using a Mach-Zehnder interferometer. Figure 4 shows the schematics of the interferometer. The interferometer consists of a HeNe laser with wavelength of 633 nm and 10 mm diameter, two 50/50 beam splitters (BS1, BS2), two mirrors (M1, M2) and a CCD camera connected to a computer. The *SlitCell* is mounted on a motorized linear stage and installed in a vacuum chamber. The laser beam of one interferometer arm passes through the windows of the vacuum chamber and the *SlitCell* lateral windows. The gas cell has been translated perpendicularly to the laser beam axis to scan the entire length of the gas cell. The translation is made in steps of 8 mm so that 2 mm region is overlapped in the scanning to insure the continuity between each step.

In order to obtain the gas density profile, we recorded two interferograms for each step: one without gas for reference and one with gas for signal. For the measurement, Argon gas was chosen due to its high refractive index n = 1.000281. The gas density can be obtained from the fringe shift of the interferogram by the formula[19] $\rho = \frac{\Delta m \, \lambda \, \rho_0}{L(n_{\text{gas}}-1)}$, where Δm is the ratio between the fringe shift with gas from the reference and the fringe distance in the reference, λ is the laser wavelength, $\rho_0 = 2.68 \times 10^{19}$ cm$^{-3}$ for Argon, *L (=3 cm)* is the lateral length of medium along the beam propagation, and $n_{gas}$ is the refractive index of the gas at atmospheric pressure with 300-K temperature. The distance between two consecutive fringes in the reference is 25 pixels which corresponds to approximately 0.13 mm spatial resolution. The noise level is about ±1 pixels in fringe shift which corresponds to a density noise of $\pm 8.54 \times 10^{16}$ atoms/cm$^3$. Since the mechanical vibrations from the environment have been present during the measurements although a stabilized optical table has been used, we have applied a relatively long exposure time of 1.5 seconds to the camera for averaging the effects of mechanical vibrations. The gas has been flown continuously in the *SlitCell* to ensure the steady state flow condition during the measurements, and, thus, the long exposure time of the camera has not blurred the fringes due to transient evolution of the gas flow.

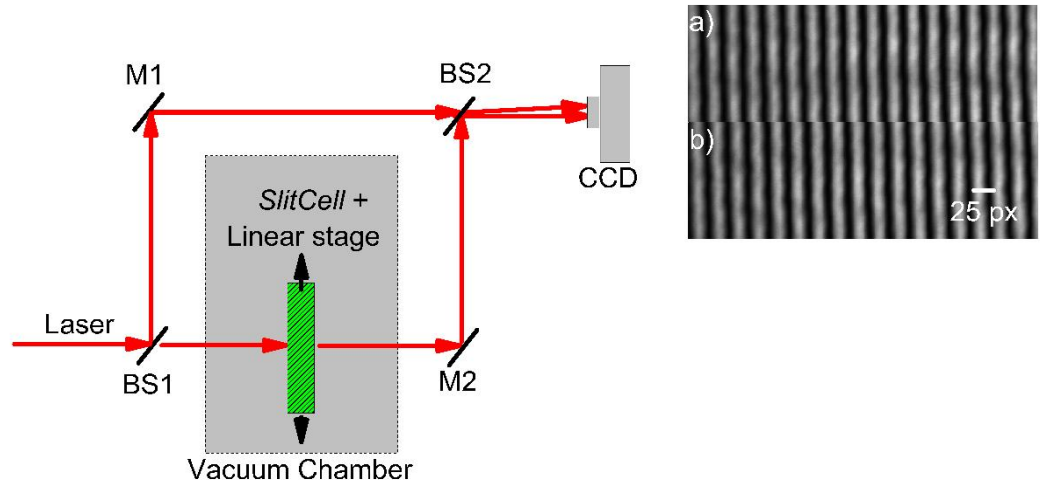

**Fig. 5 Schematic drawing of the Mach-Zehnder interferometer with the gas cell placed in one of the arms. Typical interferograms obtained a) with gas present in the *SlitCell* and b) without gas (reference). The *SlitCell* is translated along the perpendicular direction relative to the laser beam propagation.**

## 4.  COMPARISON BETWEEN MEASUREMENTS AND CFD SIMULATIONS

The *SlitCell* has been characterized first through fluid dynamics simulations, which have been validated by interferometry measurements. Figure 6 shows the measured density profiles (full line curves) along the central axis of the gas cell (straight line between the two gas outlets in the gas cell part) and CFD simulation results (dashed line) for backing pressures of 10 bar, 16 bar and 24 bar. Considering the experimental errors and ambiguities, the measured density profile closely matches the CFD simulations. There have been small differences of 6.6 % in average between measurements and simulations for the backing pressures of 16 bars and 24 bars with 80-mm long *SlitCell*, which may be caused by small geometrical differences between the manufactured and simulated *SlitCell*. In the measurement, the uniformity of density profiles along the central axis of the cell part has been in the noise level of the interferometer and has not presented any density spikes at the middle of the gas cell. Thus, we can conclude that our design of *Slitcell* has been validated as a tool for providing long and uniform gas medium for LWFA.

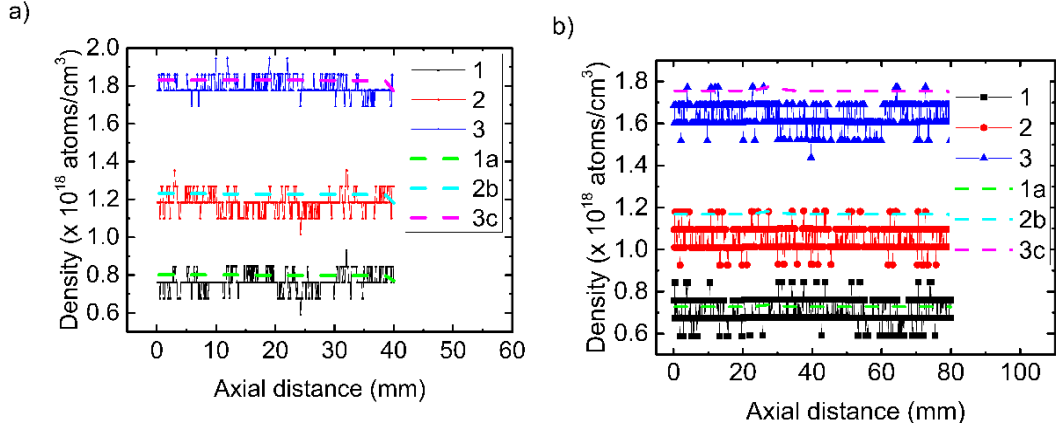

**Fig. 4 a) Gas density profile for the *SlitCell* with 40 mm length and b) gas density profile for the *SlitCell* with 80 mm length. 1a, 2b, and 3c dashed lines represent the results from simulations with Helium at 10 bar, 16 bar and 24 bar, respectively, and the 1, 2 and 3 full lines represent the results from interferometry with Argon at 10 bar, 16 bar and 24 bar respectively.**

The *SlitCell*, although simple in design, can be customized to generate much more complicated profiles. For instance, the density profiles can be shaped in step-like form as required[20] for staged LWFA[7] by installing additional pinholes with various diameters along the gas cell. Mounting each pinhole in an independent movable system could permit the control of all the parameters of the gas density profile individually in each stage. With this multi-sectional *Slitcell*, we could control the gas density profile for each section, which will allow to manipulate the parameters of accelerated electron beam such as energy, energy spread or charge. This kind of customizable *SlitCell* is currently under development and its details will be presented later in separate publication. Additional simulations (not presented here) showed that the *SlitCell* length can be extended even to several tens of centimeters without affecting the uniformity of the density profile proving once more that the *SlitCell* is a versatile gas target that could be an essential gas medium for studying various interactions between intense laser pulses and gaseous medium, especially for LWFA with present and future PW lasers[21,22].

## III.  CONCLUSIONS

We have presented a novel gas target designed for LWFA experiments. The new gas target consists of a gas cell that has a gas inlet in the shape of the slit nozzle and generates very uniform, length and density variable density profile. Although we have presented only a few experimental density profiles with minimum density of $\approx 10^{17}$ atoms/cm$^3$, the fluid dynamics simulation show that the density can be reduced as low as $10^{16}$ atoms/cm$^3$ without degrading its uniformity. This result, combined with the customizability of density profiles, gives us a versatile apparatus that could improve the present wakefield accelerators and further develop the field of laser based accelerators.

This work has been supported by the Institute for Basic Science of Korea under IBS-R012-D1.